\documentclass[preprint,12pt]{elsarticle}

\usepackage{amssymb}

\usepackage{latexsym}
\usepackage{amsmath}
\usepackage{amsfonts}
\usepackage{amsthm,tikz,mathtools}
\usepackage{psfrag}
\usepackage{graphicx}
\usepackage{algorithmic,algorithm}
\usepackage{multirow,multicol}
\usepackage{adjustbox}

\newcommand{\balpha}{\boldsymbol{\alpha}}
\newcommand{\bbeta}{\boldsymbol{\beta}}

\newcommand{\bepsilon}{\boldsymbol{\epsilon}}

\newcommand{\btheta}{\boldsymbol{\theta}}

\newcommand{\bmu}{\boldsymbol{\mu}}

\newcommand{\bxi}{\boldsymbol{\xi}}
\newcommand{\0}{\boldsymbol{0}}

\newcommand{\brho}{\boldsymbol{\rho}}

\newcommand{\bSigma}{\boldsymbol{\Sigma}}

\newcommand{\bOmega}{\boldsymbol{\Omega}}

\newcommand{\bB}{\boldsymbol{B}}
\newcommand{\bb}{\boldsymbol{b}}
\newcommand{\bA}{\boldsymbol{A}}

\newcommand{\bD}{\boldsymbol{D}}

\newcommand{\bE}{\boldsymbol{E}}

\newcommand{\bg}{\boldsymbol{g}}

\newcommand{\bH}{\boldsymbol{H}}

\newcommand{\bI}{\boldsymbol{I}}

\newcommand{\bM}{\boldsymbol{M}}

\newcommand{\bP}{\boldsymbol{P}}

\newcommand{\bR}{\boldsymbol{R}}

\newcommand{\bS}{\boldsymbol{S}}

\newcommand{\bu}{\boldsymbol{u}}
\newcommand{\bU}{\boldsymbol{U}}

\newcommand{\bV}{\boldsymbol{V}}

\newcommand{\bW}{\boldsymbol{W}}
\newcommand{\bx}{\boldsymbol{x}}
\newcommand{\bX}{\boldsymbol{X}}
\newcommand{\by}{\boldsymbol{y}}

\newcommand{\bz}{\boldsymbol{z}}
\newcommand{\bZ}{\boldsymbol{Z}}

\def\diag{\mathop{\rm diag}\nolimits}

\def\var{\mathop{\rm Var}\nolimits}

\def\corr{\mathop{\rm corr}\nolimits}
\def\E{\mathop{\rm E}\nolimits}
\def\P{\mathop{\rm P}\nolimits}
\def\logit{\mathop{\rm logit}\nolimits}

\newtheorem{thm}{Theorem}

\begin{document}

\begin{frontmatter}



\title{Semiparametric Mixed-effects Model for Longitudinal Data with Non-normal Errors}


\author{Mozhgan Taavoni\fnref{fn1}}
\author{Mohammad Arashi\corref{cor1}\fnref{fn1}}
\ead{arashi@um.ac.ir}
\cortext[cor1]{Corresponding author}
\fntext[fn1]{Department of Statistics, Faculty of Mathematical Sciences, Ferdowsi University of Mashhad, IRAN.}

\begin{abstract}
Difficulties may arise when analyzing longitudinal data using mixed-effects models if there are nonparametric functions present in the linear predictor component. This study extends the use of semiparametric mixed-effects modeling in cases when the response variable does not always follow a normal distribution and the nonparametric component is structured as an additive model. A novel approach is proposed to identify significant linear and non-linear components using a double-penalized generalized estimating equation with two penalty terms. Furthermore, the iterative approach provided intends to enhance the efficiency of estimating regression coefficients by incorporating the calculation of the working covariance matrix. The oracle properties of the resulting estimators are established under certain regularity conditions, where the dimensions of both the parametric and nonparametric components increase as the sample size grows. We perform numerical studies to demonstrate the efficacy of our proposal.
\end{abstract}




\end{frontmatter}


\section{Introduction}\label{sec1}
We often involve longitudinal studies in epidemiology, social science, and other biomedical research areas, where repeated measurements from the same subject are correlated. Under the likelihood principle, one may use the generalized linear mixed-effects models (GLMMs) to analyze longitudinal data, (\citet{Zeger:Karim:1991}, \citet{Breslow:Clayton:1993}), where subject-specific random effects explicitly take care of the correlation induced by among-subject variation. A standard GLMM assumes parametric fixed effects that may be too restrictive when some functional covariate effects are present. Comprehensive studies that cover the methodologies, computational strategies and applications of the mixed-effects models for longitudinal data can be found in \citet{Wang:Fan:2011}, \citet{Groll:Tutz:2014}, \citet{Wang:2017}, \citet{Wang:et:al:2018}, \citet{Mattos:et:al:2021}, \citet{Emmenegger:Buhlmann:2022}, and \citet{Lin:Wang:2022}. Moreover, with sparse data such as binary data, it is often not even possible to determine empirically the underlying functional form. To allow richer and more flexible model structures, an effective semiparametric regression tool is the generalized additive mixed-effects models (GAMMs) introduced by \citet{Lin:Zhang:1999}, which a linear predictor involves smooth functions of covariates. Statistical inference of additive models with longitudinal data has also been considered by some authors. This line of work includes \citet{Berhane:Tibshirani:1998}, \citet{You:Zhou:2007}, \citet{Carroll:et:al:2009}, \citet{Xue:et:al:2010}, \citet{Roozbeh:2016}, and \citet{Roozbeh:2018a}.
However, when the number of covariates is very large, selection of important variables is still a challenging issue. The literature on variable selection for high dimensional longitudinal data is rather limited due to the challenges imposed by incorporating the intracluster correlation. Some developments along this line can be found in \citet{Wang:et:al:2008}, \citet{Ni:et:al:2010}, and \citet{Ma:et:al:2013} for continuous longitudinal data, and \citet{Fu:2003} and \citet{Dziak:2006} for discrete longitudinal data. In all aforementioned studies, the dimension of predictors is  fixed. Among few cases, \citet{Wang:et:al:2012} and \citet{Chu:et:al:2016} studied variable selection of the longitudinal GLMs with a diverging number of parameters. We also refer to \citet{Roozbeh:2018} and \citet{Roozbeh:et:al:2022}. The SCAD-penalized selection procedures were illustrated in \citet{Xue:et:al:2010} for the generalized additive model with correlated data. To the best of our knowledge, simultaneous estimation and variable selection in the high-dimensional GAMMs have not been investigated, especially when the number of both linear predictors and functional components diverges with the sample size. We try to fill this gap by allowing for non-Gaussian data and nonlinear link functions.

Contrary to the \citet{Xue:et:al:2010} where only consider the additive non parametric component, modeling strategy of our work is rather different because our model uses linear part and additive nonparametric functions to combine underlying covariate effects while accounting for overdispersion and correlation by adding random effects. We consider the case where the number of linear predictors along with additive nonparametric functions are allowed to increase with the sample size under a high-dimensional regime. We give a regularization estimation procedure based on double penalization method while the additive nonparametric functions are aproximated by smoothing splines, and incorporate the within-cluster correlation to obtain an efficient parameter estimation. We apply penalty functions to the estimating equation objective function  to select the correct generalized additive model. In our method, we include additive nonparametric components in the model and use double penlization. Our numerical studies show that our proposal performs well in estimation efficiency and model selection when the dimension of linear predictors and functional components are high. 

\subsection{Plan of the paper}
The rest of paper is organized as follows. Section 2, introduces the modeling framework with the necessary notations, and spline approximation of the nonparametric function. A double penalization procedure and the choice of the tuning parameters included in Section 3. A classical SCAD penalty is used for the linear part and the nonparametric functional forms are selected groupwise for each functional component. In addition, the proposed penalized estimation involves the specification of the posterior distribution of the random effects, which cannot be evaluated in a closed form. However, it is possible to approximate this posterior distribution by producing random draws from a distribution using the Metropolis algorithm, which does not require the specification of the posterior distribution. In the same section, we establish the asymptotic theory for the proposed method in a high-dimensional framework, and achieve the oracle properties. To estimate the parameters, a computationally flexible iterative algorithm is developed. In Section 4, we apply a number of simulations to assess the finite sample performance of the proposed estimation method in the GAMM. Real data analysis is also presented in this section to augment the theoretical results. Some concluding remarks are given in Section 5. Proofs of the main results as well as some instrumental lemmas are provided in a separate supplementary file.

\section{Semiparametric Mixed-effects Model and Approximation}
Consider a longitudinal study with $n$ subjects and $n_i$ observations over time for each subject with $i=1,\ldots,n$. Let $\bu_i$ be a $q\times 1$ vector of random effects corresponding to
the $i$th subject, $y_{ij}$, $j = 1,\ldots, n_i$, denote the longitudinal response for subject $i$ measured at time $t_{ij}$, $\bx_{ij}=(x_{ij1},\ldots, x_{ijp_n})^\top$ is a $p_n\times 1$ vector of scalar covariates, $\bz_{ij}=(z_{ij1},\ldots, z_{ijq})^\top$ is
a $q\times 1$ vector of explanatory variables relating to the random effects. Suppose that $\by_i=(y_{i1}, \ldots, y_{in_i})^\top$ given $\bu_i$ are conditionally independent and each $y_{ij}|\bb_i$ is distributed according to an exponential dispersion model with mean $\mu_{ij}=\E(y_{ij}|\bb_i)$ and conditional variance $\var(y_{ij}|\bu_i)=\phi \omega^{-1}_{ij}\nu(\mu_{ij})$, where $\nu(.)$ is a specified variance function, $\omega_{ij}$ is a known weight (e.g., a binomial denominator), and $\phi$ is a dispersion parameter, which can be either known or required to be estimated. In addition to random and fixed effects, the model includes an additive term $\bg_{i}(x_{ijk})=\lbrace g_{i1}(x_{ij1}), \ldots, g_{ir_n}(x_{ijr_n})\rbrace$, a $r_n\times 1$
vector of functional predictors. The $g_{ik}(x_{ijk})$ are
unknown smooth but arbitrary continuous and twice differentiable functions of covariate $x_{ijk}$, on the bounded 
and closed interval $j_g$, with unknown smooth mean functions $\E\big\lbrace g_{ik}(x_{ijk})\big\rbrace$ and eigenvalue, eigenfunction pairs $\lbrace\pi_{kl}, \phi_{kl}(x_{ijk})\rbrace_{l \geq 1}$, $k=1,\ldots,r_n$. Then a GAMM is of the form
\begin{eqnarray}\label{eq1}
g(\mu_{ij})=\bx_{ij}^\top\bbeta_n+\sum_{k=1}^{r_n}g_{ik}(x_{ijk})+\bz_{ij}^\top\bu_i,
\end{eqnarray}
where $g(.)$ is a known monotonic and differential link function, $\bx_{ij}^\top\bbeta_n$ is a linear parametric term with parameter vector $\bbeta_n=(\beta_1, \ldots, \beta_{p_n})^\top$, $\sum_{k=1}^{r_n}g_{ik}(x_{ijk})$ is an
additive term with unspecified influence functions $g_{i1},\ldots,g_{ir_n}$ and finally $\bz_{ij}^\top\bu_i$ contains the random effects part. The $q$-dimensional random effects $\bu=\lbrace \bu_1,\ldots, \bu_q \rbrace$ are assumed to be independently  and identically distributed, depending on parameters $\bSigma$ as $\bu_i \sim f_u(\bu_i|\bSigma)$. If the functions $g_k(.)$ are strictly linear, the model reduces to the common GLMM. Note that the dimension of random effects $q$ is assumed to be fixed, while the dimension of both scalar covariates and functional predictors $p_n$ and $r_n$ may grow to infinity with the sample size $n$. Specific assumptions will be considered for the number of variables $p_n$ and $r_n$ in section 3.3. 

In the estimation part, we approximate the smooth functions $\lbrace g_{ik}(.)\rbrace_{k=1}^{r_n}$ by polynomial splines, which has been used by \citet{Xue:Liang:2009}, \citet{Xue:et:al:2010}, \citet{Groll:Tutz:2012}, and \citet{Wu:Xue:2014}. For each $k=1,\ldots,p_n$, let $\nu_{k}= \lbrace 0= x_{k,0}< x_{k,1}< \ldots< x_{k,L_{nk}}< x_{k,L_{nk}+1} =1\rbrace$ be a partition of the interval $\lbrack 0, 1 \rbrack$, with $L_{nk}$ interior knots. The polynomial splines of order $d+1$ are functions with $d$-degree (or less) of polynomials on intervals $\lbrack x_{k,k^{'}}, x_{k,k^{'}+1})$, $k^{'}=0,\ldots, L_{nk}-1$, and $\lbrack x_{k,L_{nk}}, x_{k,L_{nk}+1}\rbrack$, and have $p-1$ continuous derivatives globally. We denote the space of such spline functions by $\phi_k=\big(\phi^p\lbrack 0,1 \rbrack, \nu_k\big)$. Denote $\phi_k^0=\lbrace g \in \phi_k:\int_{0}^{1} g(x)dx=0 \rbrace$, which consists of centered spline functions. Let 
$\lbrace B_{kl}(.)\rbrace_{l=1}^{h_{nk}}$ be a set of spline bases of $\phi_k^0$ for variable $k$ with $h_{nk}=L_{nk}+d$. Here, the centered truncated power basis is used in implementation, with $\big\lbrace B_{kl}(.)=b_{kl}-\E_n(b_{kl})\big\rbrace_{l=1}^{h_{nk}}$, where $\bb_k=\lbrace b_{k1},\ldots,b_{kh_{nk}}\rbrace$ is the truncated power basis given as $\lbrace x_k,\ldots,x_k^d,(x_k-x_{k,1})_+^d,\ldots,(x_k-x_{k,L_{nk}})_+^d \rbrace$, in which $(x)_+=\max(0,x)$. Then each coefficient function $g_{ik}(.)$ for $k=1,\ldots,r_n$ can be approximated by
$g_{ik}(x_{ijk})\thicksim\sum_{l=1}^{h_{nk}} \alpha_{kl}B_{kl}(x_{ijk})$, where levels of is the spline coefficients. The subscript $k$ indicates that different coefficient functions have different smoothness. Let $\balpha_{nk}=(\alpha_{k1},\ldots,\alpha_{kh_{nk}})^\top$ denote the unknown parameter vector of the $k$th smooth function and $\bB_{k}(x_{ijk})=\big(B_{k1}(x_{ijk}), \ldots,B_{kh_k}(x_{ijk})\big)^\top$ represent the vector-valued evaluations of the $h_{nk}$ basis functions. Then, model \eqref{eq1} has the parameterized form:
\begin{eqnarray*}
	g(\mu_{ij})=\bx_{ij}^\top\bbeta_n+ \bB_{1}^\top(x_{ij1})\balpha_{n1}+\ldots+\bB_{r_n}^\top(x_{ijr_n})\balpha_{nr_n}+\bz_{ij}^\top\bu_i.
\end{eqnarray*}
By collecting observations within one subject one obtains the design matrix $\bX_i=(\bx_{i1},\ldots, \bx_{in_i})^\top$ for the $i$th covariate, and analogously we set $\bZ_i=(\bz_{i1},\ldots, \bz_{in_i})^\top$, so that the model has the simpler form
\begin{eqnarray*}
	g(\bmu_{i})=\bX_{i}\bbeta_n+ \bB_{i1}\balpha_{n1}+\ldots+\bB_{ir_n}\balpha_{nr_n}+\bZ_{i}\bu_i,
\end{eqnarray*}
where $\bB_{ik}=\big(\bB^\top_{1}(x_{i1k}),\ldots,\bB^\top_{k}(x_{in_ik})\big)^\top$ denotes the spline design matrix of the $i$th
subject and variable $k$ and $g$ is understood componentwise. Furthermore, let $\bX=(\bX_1^\top,\ldots,\bX_n^\top)^\top$, 
$\bZ= \diag(\bZ_1,\ldots,\bZ_n)$ be a block-diagonal matrix and, $\bu=(\bu_1^\top,\ldots,\bu_n^\top)^\top$
be the vector collecting all random effects. Then, it yields
\begin{eqnarray}\label{eq2}
g(\bmu)=\bX\bbeta_n+ \bB_{1}\balpha_{n1}+\ldots+\bB_{r_n}\balpha_{nr_n}+\bZ\bu,
\end{eqnarray}
where $\bB_k=(\bB_{1k}^\top,\ldots,\bB_{nk}^\top)^\top$ represens the spline design matrix of the $k$th smooth function. The model can be more simplified to 
\begin{eqnarray*}
	g(\bmu)=\bX\bbeta_n+\bB\balpha_n+\bZ\bu,
\end{eqnarray*}
where $\balpha_n=(\balpha_{n1}^\top,\ldots,\balpha_{nr_n}^\top)^\top$ and $\bB=[\bB_1^\top, \ldots,\bB_{r_n}^\top]^\top$. An alternative form that we also use in the following is
$$ g(\bmu)= \bD\btheta_n+\bZ\bu,$$
where $\bD=(\bX^\top,\bB^\top)^\top$ and $\btheta_n=(\bbeta_n^\top,\balpha_n^\top)^\top$.

\section{Regularization in the Semiparametric Model}

Focusing on the GLMM,we assume that the conditional density of $y_{ij}|\bu_i$, belongs to the exponential family whose probability density function (p.d.f) is given by
\begin{eqnarray*}
	p(y_{ij}|\bu_i)= \exp[\phi^{-1}\lbrace y_{ij}\theta_{ij} -b(\theta_{ij})\rbrace+ c(y_{ij},\phi)],
\end{eqnarray*}
where $\theta_{ij}$ is the (scalar) canonical parameter, $b(\theta_{ij})$ is a specific function corresponding to the type of exponential family, $c(.)$ the log normalization constant and $\phi$ the dispersion parameter. Then the quasi-likelihood (QL) of the model parameters $(\btheta_n,\bSigma,\phi)$ can be expressed as
\begin{eqnarray}\label{eq3}
L(\btheta_n, \boldsymbol{\Sigma}, \phi)=\prod_{i=1}^n\int p_{\by_{i}|\bu_i}(\by_{i}|\bu_i, \btheta_n, \phi) p_{\bu}(\bu_i|\boldsymbol{\Sigma})d\bu_i,
\end{eqnarray}
where $p_{\by_{i}|\bu_i}(\by_{i}|\bu_i, \btheta_n, \phi)=\prod_{j=1}^{n_i} p(\by_{ij}|\bu_i, \btheta_n, \phi)$. 
In what follows, we consider the selection and estimation of both functional and scalar parameters.  Identifying important variables is a crucial step in analyzing high-dimensional data, because each redundant variable involves an infinite dimension of parameters for nonparametric components. Here a predictor variable $x$ is said to be redundant in
model \eqref{eq1}, if and only if $x=0$ or $g(x)=0$ almost surely. Otherwise, a predictor variable $x$ is said to be relevant. Suppose that the true value $\bbeta_{n0}$ of the regression coefficient $\bbeta_n$ can be decomposed into $\bbeta_{n0}=(\bbeta_{n01}^\top,\bbeta_{n02}^\top)^\top$
and the corresponding design matrix into $\bX_i = (\bX_{i(1)},\bX_{i(2)})$, where $\bbeta_{n01}$ is a $q_n\times 1$ vector corresponding to significant covariate effects $\bX_{i(1)}$ and $\bbeta_{n02}$ is a $(p_n - q_n)\times1$ vector of zeros. Also assume that only the first $d_n$ functional predictors are significant or, equivalently that the true values of $g_{ik}(x_{ijk})$ are such that $g_{ik}(x_{ijk})\equiv 0$ for $k= d_n + 1, \ldots , r_n; x_{ijk} \in J_x$. In other words, $\balpha_{n(d_n+1)}, \ldots, \balpha_{nr_n}$ are zero vectors. Our goal is to consistently identify such subsets of relevant variables and estimate their unknown parameters or function components. In order to address the difficulty caused by the infinite dimensionality of functional predictors and high dimensionality of scalar covariates, the terms $g_{ik}(x_{ijk})$ in \ref{eq1} are first approximated with truncated linear predictors $\sum_{l=1}^{h_{nk}} \alpha_{kl}B_{kl}(x_{ijk})$ for some large numbers $\lbrace h_{nk}\rbrace_{k=1}^{r_n}$. The numbers $h_{nk}$ may vary with the sample size $n$, and play the role of smoothing parameters which balance the trade-off between bias and variance. Then, a suitable regularization is imposed on functional predictor $\bg_i(.)$ and scalar covariate $\bX$, respectively. As a consequence, the coefficient functions of unimportant functional predictors and regression parameters of unimportant scalar covariates will be shrunken to zero. Therefore, the goals of the variable selection and estimation of important functional coefficients and fixed-effects regressor coefficients can be achieved by maximizing the penalized QL
\begin{eqnarray}\label{eq4}
\ell(\btheta_n, \bSigma, \phi)&=&\sum_{i=1}^n \hbox{ln} p_{\by_{i}|\bu_i}(\by_{i}|\bu_i, \btheta_n, \phi)+\sum_{i=1}^n \hbox{ln} p_{\bu_i}(\bu_i|\bSigma)\\ \nonumber
&-&n\sum_{k=1}^{p_n}p_{\lambda_{n}}(|\beta_{nk}|)-n\sum_{k=1}^{r_n}p_{\lambda_{nk}}(\Vert\balpha_{nk}\Vert_{\bW_k}),
\end{eqnarray}
where $p_{\lambda_{n}}(.)$ and $p_{\lambda_{nk}}(.)$ are penalty functions with tuning parameters $\lambda_{n}$, $\lambda_{nk}$, and $\Vert\balpha_{nk}\Vert_{\bW_k}=\sqrt{\balpha_{nk}^\top \bW_k\balpha_{nk}}$ with $\bW_k=\frac{1}{n}\sum_{i=1}^n\frac{1}{n_i}\sum_{j=1}^{n_i}\bB_k(\bx_{ijk})\bB^\top_k(\bx_{ijk})$, the weighted Euclidean norm associated with a group penalty. In order to regularize all predictors on a comparable scale, one often standardizes the predictors before imposing a penalty on the objective function. Thus, we standardize $(\bx_{1k},\ldots, \bx_{nk})$ to have unit variance, for $k=1,\ldots, p_n$. Since the variability of the $k$th functional predictor $g_{ik}(x_{ijk})$ can be approximated by $\sum_{l=1}^{h_{nk}} \widehat{\pi}_{kl}$, and standardization of $g_{ik}(x_{ijk})$ is equivalent to adding weight to the penalty function, we suggest using $\lambda_{nk}= \lambda_{n} \Big(\sum_{l=1}^{h_{nk}} \widehat{\pi}_{kl}\Big)^{1/2}$, which simplifies both the computation and theory. The resulting estimators for $\beta_{nk}$ and $\balpha_{nk}$ obtained from \eqref{eq4} are then denoted by $\widehat{\beta}_{nk}$ and $\widehat{\balpha}_{nk}$.
For the maximum QL (MQL) estimates of the parameters $(\btheta_n,\bSigma,\phi)$, one can maximize the function \eqref{eq4} by using suitable numerical techniques. 

The EM algorithm is an attractive method to obtain the ML estimates, in presence of incomplete data, which avoids explicit calculation of the observed data log-likelihood. To set up the EM algorithm we consider the random effects, $\bu_i$, to be the missing data and the complete data, is then $(\by_{i},\bu_i)$. Considering $\bu_i$ to be the missing has an advantage that in the M-step, maximization can be accomplished with respect to the parameters $\btheta_n$ and $\phi$ in the first, third, and fourth terms of \eqref{eq4}. Thus, the M-step with respect to $(\btheta_n,\phi)$ uses only the GLM part of the penalized likelihood function. Hence, the procedure is similar to a standard GLM computation assuming $\bu_i$ is known. Therefore, maximizing with respect to $\bSigma$, in the second term, can be handled by the ML using the distribution of $p_{\bu_i}(\bu_i|\bSigma)$ after replacing sufficient statistics with the conditional expected values. It thus makes sense to develop a simulation analogous to the Newton-Raphson approach for fitting the GAMM. Using this separation as in \eqref{eq4} and the fact that 
$\frac{\partial p_{\lambda}(|\beta_{nk}|)}{\partial\beta_{nk}}= p^{'}_{\lambda}(|\beta_{nk}|)\hbox{sign}(\beta_{nk})$ and 
$\frac{\partial p_{\lambda}(\Vert\balpha_{nk}\Vert_{\bW_k})}{\partial\balpha_{nk}}=p^{'}_{\lambda}(\Vert\balpha_{nk}\Vert_{\bW_k})\Vert\balpha_{nk}\Vert_{\bW_k}^{-1}\bW_k\balpha_{nk}$, 
the penalized MQL (PMQL) equations for $\btheta_n$ and $\bSigma$ take the following forms
\begin{eqnarray*}
	&&\E\Big[\frac{\partial \ln p_{\by_{i}|\bu_i}(\by_{i}|\btheta_n,\phi)}{\partial \btheta_n} \Big|y_{ij}\Big]-\sum_{k=1}^{p_n}q_{\lambda_n}(|\bbeta_n|)\hbox{sign}(\bbeta_n)-\sum_{k=1}^{p_n}q_{\lambda_{nk}}(\Vert \balpha_{nk}\Vert_{\bW_k})\Vert \balpha_{nk}\Vert_{\bW_k}^{-1}=\0;\\ \nonumber
	&&\E\Big[\frac{\partial \ln p_{\bu_i}(\bu_{i}|\bSigma)}{\partial \bSigma}|y_{ij}\Big]=\0,
\end{eqnarray*}
with $\hbox{sign}(a) = I(a> 0) - I(a < 0)$ and $q_{\lambda}(.)$ is the first-order derivative of $p_{\lambda}(.)$. The PML estimates of $(\bbeta_n,\balpha_n)$ and $\bSigma$ can be obtained by solving the preceding equations numerically. The optimal penalized generalized estimating equation (PGEE) for $\btheta_n$ is given by
\begin{eqnarray}\label{eq5}
&&\E_{\bu|\by} \Big[n^{-1}\sum_{i=1}^n\frac{\partial \bmu_i(\btheta_n, \bu_i)}{\partial \btheta_n^\top}\bV_i^{-1}(\btheta_n, \bu_i)\big(\by_i-\bmu_i(\btheta_n, \bu_i)\big)\Big]\\ \nonumber 
&&-\sum_{k=1}^{p_n}q_{\lambda_n}(|\bbeta_n|)\hbox{sign}(\bbeta_n)-\sum_{k=1}^{p_n}q_{\lambda_{nk}}(\Vert \balpha_{nk}\Vert_{\bW_k})\Vert \balpha_{nk}\Vert_{\bW_k}^{-1}=\0,
\end{eqnarray}
where $\bmu_i(\btheta_n, \bu_i)=(\mu_{i1},\ldots,\mu_{in_i})^\top$ and $\bV_i(\btheta_n, \bu_i)$ is the covariance matrix of $\by_{i}|\bu_i$. In real applications the true intracluster covariance structure is often unknown. We assume the working correlation matrix 
$\bR(\brho): \bV_i(\btheta_n, \bu_i) = \bA_i^{\frac{1}{2}}(\btheta_n, \bu_i )\bR(\brho)\bA_i^{\frac{1}{2}}(\btheta_n , \bu_i )$, where $\brho$ is a finite dimensional parameter and $\bA_i(\btheta_n , \bu_i )=\diag(\nu_{i1},\ldots,  \nu_{in_i})$. Some commonly used working correlation structures include independence, autocorrelation (AR)-1, exchangeable, toeplitz, M-dependent, or unstructured correlation, among others. For a given working correlation structure, $\brho$ can be estimated using the residual-based method of moments. With the estimated working correlation matrix $\widehat{\bR}\equiv \bR(\widehat{\brho}) $, the PGEE in \eqref{eq5} reduces to
\begin{eqnarray*}
	\bU_n(\btheta_n)=\bS_n(\btheta_n)-\sum_{k=1}^{p_n}q_{\lambda_n}(|\bbeta_n|)\hbox{sign}(\bbeta_n)-\sum_{k=1}^{p_n}q_{\lambda_{nk}}(\Vert \balpha_{nk}\Vert_{\bW_k})\Vert \balpha_{nk}\Vert_{\bW_k}^{-1}\bW_k\balpha_{nk}=\0,
\end{eqnarray*}
where $\bS_n(\btheta_n)=\E_{u|y} \Big[\sum_{i=1}^n \bD_i^\top  \bA_i^{\frac{1}{2}}(\btheta_n , \bu_i ) \widehat{\bR}^{-1} \bA_i^{-\frac{1}{2}}(\btheta_n , \bu_i )\big(\by_i-\bmu_i(\btheta_n , \bu_i)\big)\Big]$.
We formally define the estimator as the solution $\widehat{\btheta}_n=(\widehat{\bbeta}^\top,\widehat{\balpha}^\top)^\top$ of the above estimating equations. 

Among all penalty functions, the smoothing clipped absolute deviation (SCAD) penalty proposed by \citet{Fan:Li:2001} 
can be used to retain the good features of both subset selection and ridge regression, for producing sparse solutions, and to ensure continuity of the selected models. Therefore, we will use the SCAD penalty in our simulation and application studies. The SCAD penalty function is defined by
$$q_{\lambda}(\theta) =p^{'}_{\lambda_n}(\theta)=\lambda\Big\lbrace  I(\theta\leq \lambda)+\frac{(a\lambda-\theta)_+}{(a-1)\lambda}I(\theta> \lambda) \Big\rbrace; \quad a>2,$$
where the notation $(.)_+$ stands for the positive part of $(.)$.

\subsection{Computational algorithm}
Using the local quadratic approximation (LQA), in the neighborhoods of the true parameter values $\beta_{n0k}$, $|\beta_{n0k}|>0$, the derivative of the penalty function is well approximated by $q_{\lambda_n}(|\beta_{nk}|)\hbox{sign}(\beta_{nk})\thickapprox\frac{q_{\lambda_n}(|\beta_{n0k}|)}{|\beta_{n0k}|}\beta_{nk}$. If $\balpha_{nk}$ is close to $0$ in the sense that $\Vert\balpha_{nk}\Vert_{\bW_k}\leq\epsilon$, for some small threshold value $\epsilon$, set $\balpha_{nk}$ to $\0$. In the implementation, we have used $\epsilon= 10^{-6}$. \citet{Xue:2009} and \citet{Xue:et:al:2010} applied the following LQA for the nonzero functional component $\balpha_{nk}$, that for an initial value $\Vert \balpha_{n0k}\Vert_{\bW_k}>\epsilon$, one can locally approximate the penalty function by 
\begin{eqnarray*}
	p_{\lambda_{nk}}(\Vert \balpha_{nk}\Vert_{\bW_k})
	&\thickapprox& p_{\lambda_{nk}}(\Vert \balpha_{n0k}\Vert_{\bW_k})\\
	&+&\frac{1}{2}p^{'}_{\lambda_{nk}}(\Vert \balpha_{n0k}\Vert_{\bW_k})\Vert \balpha_{n0k}\Vert^{-1}_{\bW_k}(\balpha^\top_{nk}\bW_k\balpha_{nk}-\balpha^\top_{n0k}\bW_k\balpha_{n0k}).
\end{eqnarray*}
With the mentioned LQAs, we apply the Newton-Raphson method to solve $\bU_n(\widehat{\btheta}_n)= 0$, and get the following updating formula
\begin{eqnarray}\label{eq6}
\widehat{\btheta}_n^{(m+1)}=\widehat{\btheta}_n^{(m)}+
\Big \lbrace \bH_n(\widehat{\btheta}_n^{(m)})+n \bE_n(\widehat{\btheta}_n^{(m)})\Big\rbrace^{-1} \times
\Big\lbrace  \bS_n(\widehat{\btheta}_n^{(m)})+n \bE_n(\widehat{\btheta}_n^{(m)})\widehat{\btheta}_n^{(m)} \Big\rbrace,
\end{eqnarray}
where 
\begin{eqnarray*}
	\bH_n(\widehat{\btheta}_n^{(m)})=\E_{u|y} \Big[\sum_{i=1}^n \bD_i^\top  \bA_i^{\frac{1}{2}}(\btheta_n , \bu_i ) \widehat{\bR}^{-1} \bA_i^{\frac{1}{2}}(\btheta_n , \bu_i )\bD_i\Big],
\end{eqnarray*}
\begin{eqnarray*}
	\bE_n(\widehat{\btheta}_n^{(m)})=\diag\Big \lbrace \frac{q_{\lambda_n}(|\beta_{n1}|)}{\epsilon+|\beta_{n1}|},  \ldots, \frac{q_{\lambda_n}(|\beta_{np_n}|)}{\epsilon+|\beta_{np_n}|},&& q_{\lambda_{n1}}(\Vert \balpha_{n1}\Vert_{\bW_1})\Vert \balpha_{n1}\Vert_{\bW_1}^{-1},\ldots, \\ &&q_{\lambda_{nr_n}}(\Vert \balpha_{nr_n}\Vert_{\bW_{r_n}})\Vert \balpha_{nr_n}\Vert_{\bW_{r_n}}^{-1} \Big\rbrace.
\end{eqnarray*}
After obtaining the estimator $\widehat{\btheta}_n=(\widehat{\bbeta}_n^\top,\widehat{\balpha}_n^\top)^\top$ through penalization in \eqref{eq6}, for any given $\bx\in\lbrack 0,1\rbrack^{r_n}$, an estimator of the unknow functional components in \eqref{eq1} is given as
\begin{eqnarray*}
	\widehat{g}_k(x_{ijk})=\sum_{l=1}^{h_{nk}}\widehat{\balpha}_{kl}B_{kl}(x_{ijk}).
\end{eqnarray*}

Note that, in general, the expectations in \eqref{eq6} cannot be computed in a closed form as the conditional distribution of $\bu_i|\by_i$ involves the marginal distribution of $\by_i$, which is not easy to be computed explicitly. In the following we outline the computational procedure used for sample generation. 
Let $\bU$ denote the previous draw from the conditional distribution of $\bU|\by$, and generate a new value $u_k^*$ for the $j$th component of $\bU^*=(u_1, \ldots , u_{k-1}, u^*_k, u_{k+1}, \ldots, u_{nq})$ by using the candidate distribution $p_{\bu}$, accept $\bU^*$ as the new value with probability
\begin{eqnarray}\label{eq7}
\alpha_k(\bU,\bU_*)=\hbox{min}\Big\lbrace 1,  \frac{p_{\bu|\by}(\bU^*|\by,\btheta_n,\bD)p_{u}(\bU|\bD)}{p_{\bu|\by}(\bU|\by,\btheta_n,\bD)p_{u}(\bU^*|\bD)} \Big\rbrace.
\end{eqnarray}
otherwise, reject it and retain the previous value $\bU$. The second term in brace in \eqref{eq7} can be simplified to
\begin{eqnarray*}
	\frac{p_{\bu|\by}(\bU^*|\by,\btheta_n,\bD)p_{u}(\bU|\bD)}{p_{\bu|\by}(\bU|\by,\btheta_n,\bD)p_{u}(\bU^*|\bD)}&=&\frac{p_{\by|\bu}(\by|\bU^*,\btheta_n)}{f_{\by|\bu}(\by|\bU,\btheta_n)}\\ \nonumber
	&=&\frac{\prod_{i=1}^n p_{\by_i|\bu}(\by_i|\bU^*,\btheta_n)}{\prod_{i=1}^n f_{\by_i|\bu}(\by_i|\bU,\btheta_n)}.
\end{eqnarray*}
Note that, the calculation of the acceptance function $\alpha_k(\bU,\bU_*)$ here involves only the specification of the conditional distribution of $\by|\bu$ which can be computed in a closed form. 

The suggested double penalized approach can be implemented step by step, and the detailed algorithm  can be summarized in Algorithm 1 descring the Metropolis step into the Newton-Raphson iterative equation \eqref{eq6} for the Monte Carlo estimates of expected values.

\begin{algorithm}\label{al1}
	\caption{Monte Carlo Newton-Raphson (MCNR) algorithm}
	\begin{algorithmic}[]
		\item[{\it \bf step 1.}] Smooth each predictor trajectory $\lbrace g_{ik}(.)\rbrace_{k=1}^{p_n}$, by local linear smoothing technique. The smoothed predictor trajectories are then denoted as $g_{ik}(.)=\bB_{ik}\balpha_{nk}$. Set $m_k=0$. Choose initial values $\btheta_n^0=(\bbeta_n^{0\top},\balpha_n^{0\top})^\top$ and $\bSigma^0$.
		\item[{\it \bf step 2.}] Generate $N$ observations $\bU^{(1)}, \ldots , \bU^{(N)}$ from the distribution $p_{\bu|\by} (\bu|\by, \btheta_n^{(m_k)}, \bSigma^{(m_k)})$ using the Metropolis algorithm. Use these observations to find the Monte Carlo estimates of the expectations. Specially,
		\item[a)] Compute $\btheta_n^{(m_k+1)}$ from the expression
		\begin{eqnarray*}
			\btheta_n^{(m_k+1)}&=&\btheta_n^{(m_k)}+\Big \lbrace \frac{1}{N}\sum_{k=1}^N\Big[ \bH_n\big(\widehat{\btheta}_n^{(m_k)}, \bU^{(k)}\big)\Big]+n \bE_n(\widehat{\btheta}_n^{(m_k)})\Big\rbrace^{-1}\\ \nonumber
			&\qquad & \qquad \times\Big\lbrace \frac{1}{N}\sum_{k=1}^N\Big[\bS_n(\widehat{\btheta}_n^{(m_k)}, \bU^{(k)})\Big]-n\bE_{n}(\widehat{\btheta}_n^{(m_k)})\widehat{\btheta}_n^{(m_k)}\Big\rbrace,
		\end{eqnarray*}
		where
		\begin{eqnarray*}
			\bH_n\big(\widehat{\btheta}_n^{(m_k)}, \bU^{(k)}\big)&=&\sum_{i=1}^n \bD_i^\top  \bA_i^{\frac{1}{2}}(\btheta_n^{(m_k)} , U_i^{(k)}) \widehat{\bR}^{-1} \bA_i^{\frac{1}{2}}(\btheta_n^{(m_k)} , U_i^{(k)} )\bD_i,\\ \nonumber
			\bS_n\big(\widehat{\btheta}_n^{(m_k)}, \bU^{(k)}\big)&=&\sum_{i=1}^n \bD_i^\top  \bA_i^{\frac{1}{2}}(\btheta_n^{(m_k)} , U_i^{(k)}) \widehat{\bR}^{-1} \bA_i^{-\frac{1}{2}}(\btheta_n^{(m_k)} , U_i^{(k)} )\big(\by_i-\bmu_i(\btheta_n^{(m_k)} , U_i^{(k)})\big).
		\end{eqnarray*}
		\item[b)]  
		Compute $\bSigma^{(m_k+1)}$ by maximizing
		$
		\frac{1}{N}\sum_{k=1}^N \hbox{ln} f_u(\bU^{(k)}|\bSigma).
		$
		\item[c)] Set $m_k=m_k+1$.
		\item[{\it \bf step 3.}]  Go to step 2 until convergence is achieved. Choose $\btheta_n^{(m_k+1)}$ and $\bSigma^{(m_k+1)}$ to be the MCNR estimates of $\btheta_n$ and $\bSigma$.
	\end{algorithmic}
\end{algorithm}

To implement the proposed double penalized procedure \eqref{eq5}, one needs to choose appropriate spline spaces $\lbrace\phi_{k}^{n}\rbrace_{k=1}^{p_n}$ and tuning parameters $\lambda_{nk}$, $a$, and $\lambda_n$ involved in the SCAD penalty. The selection of knots is generally an important aspect of spline smoothing. In this paper, our main focus is inference on the parameter $\bbeta_n$. \cite{He:et:al:2005} found that knot selection is less critical for the estimate of $\bbeta_n$ than for the estimate of $\bg(.)$. They also pointed out that in most applications, the primary focus is inference on the parameter $\bbeta$, along with understanding some basic features of $\bg(.)$. Therefore, they are more concerned with the efficiency of the parameter estimate. For those reasons and the sake of simplicity, \cite{He:et:al:2005} opt for convenient choices of knot placement. More specifically, they use the sample quantiles of $\bx_{i}$ as knots. For example, in the case of three internal knots, these are taken to be the three quartiles of the observed $\bx_{i}$. They use cubic splines (i.e., splines of order 4) and take the number of internal knots to be the integer part of $M^{1/5}$, where $M$ is the number of distinct values in $\bx_{i}$.  For the choices of $\lbrace\phi_{k}^{n}\rbrace_{k=1}^{p_n}$ we use splines with equally spaced knots and fixed degrees, and select only $L_{n}$, the number of interior knots using the data. \citet{Qin:Zhu:2007} noted that the number of distinct knots has to increase with sample size for asymptotic consistency. On the other hand, too many knots would increase the variance of estimators. Therefore, the number of knots must be properly chosen to balance between the bias and variance. When $n$ goes to $\infty$, the number of knots should increase at $n^{1/(2r+1)}$. Consequently, in this paper, the number of internal knots is taken to be $L_n\approx  n^{1/(2r+1)}$ where $r$ is an integer, and select the internal knots equally spaced in the percentile ranks of $\bx_{i}$. The $r$ as assumed to be fixed at $r=2$, therefore we choose $L_n \approx n^{1/5}$. This particular choice is consistent with the asymptotic theory. However, it is mainly based on practical experience and a desire for simplicity, and by no means is it an optimal choice. Data-adaptive choices for the number or the placement of knots can be made using leave-one-cluster-out, but we do not pursue this direction here. To reduce the computational burden, we set $\lambda_{nk}= \lambda_{n} \Big(\sum_{l=1}^{h_{nk}} \widehat{\pi}_{kl}\Big)^{1/2}$ and
set $a=3.7$. Finally we need to choose $\lambda_n$. The tuning parameter $\lambda_n$ controls the sparsity of both the functional and the regression coefficients, then to obtain the optimal value of $\lambda_{n}$, one can modify the generalized cross validation (GCV) procedure. Due to the lack of joint likelihood in the generalized model, to select the tuning parameter $\lambda_{n}$, we use GCV as defined by
\begin{eqnarray*}
	\hbox{GCV}_{\lambda_n}=\frac{\hbox{RSS}(\lambda_n)/n}{(1-d(\lambda_n)/n)^2},
\end{eqnarray*}
where 
\begin{eqnarray}\label{eq8}
\hbox{RSS}(\lambda_n)=\frac{1}{N}\sum_{k=1}^N\Big[  \sum_{i=1}^n \big(\by_i-\bmu_i(\widehat{\btheta}_n , U_i^{(k)})\big)^\top\bW_i^{-1}\big(\by_i-\bmu_i(\widehat{\btheta}_n , U_i^{(k)})\big) \Big]
\end{eqnarray} 
is the residual sum of squares, and 
\begin{eqnarray*}
	d(\lambda_n)=tr\Big[ \Big \lbrace \frac{1}{N}\sum_{k=1}^N\Big[ \bH_n\big(\widehat{\btheta}_n, \bU^{(k)}\big)\Big]+n \bE_n(\widehat{\btheta}_n)\Big\rbrace^{-1}\times \Big \lbrace \frac{1}{N}\sum_{k=1}^N\Big[ \bH_n\big(\widehat{\btheta}_n, \bU^{(k)}\big)\Big])\Big\rbrace \Big]
\end{eqnarray*} 
is the effective number of parameters. Then, $\lambda_{opt}$ is the minimizer of the $\hbox{GCV}_{\lambda_n}$.
Note that $\bW_i$ in \eqref{eq8} is an $n_i\times n_i$ covariance matrix of $\by_i$, that can be computed as
$
\bW_i=\E_{u|y}  \Big(\hbox{var}(\by_i|\bu_i)\Big)+\hbox{var}_{u|y}  \Big(\E(\by_i|\bu_i)\Big),
$
where
\begin{eqnarray*}
	\E_{u|y}  \Big(\hbox{var}(\by_i|\bu_i)\Big)&=&\frac{1}{N}\sum_{k=1}^N\Big[ \bV_i(\widehat{\btheta}_n, U_i^{(k)}) \Big],\\
	\var_{u|y}  \Big(\E(\by_i|\bu_i)\Big)&=&\frac{1}{N}\sum_{k=1}^N\Big[\bmu_i(\widehat{\btheta}_n , U_i^{(k)})\Big]^2-\bigg[ \frac{1}{N}\sum_{k=1}^N\Big[\bmu_i(\widehat{\btheta}_n , U_i^{(k)})\Big]\bigg]^2.
\end{eqnarray*}

\subsection{Oracle properties}\label{sec asymp}
Our proposed estimator for $\btheta_n$ is the solution of $\bU_n(\btheta_n)=\0$. Because $\bU_n(\btheta_n)$ has discontinuous points, an exact solution to $\bU_n(\btheta_n)=\0$ may not exist. We formally define the estimator $\widehat{\btheta}_n$ to be an approximate solution, i.e., $\bU_n(\widehat{\btheta}_n)= o(a_n )$ for a sequence $a_n \to 0$. The rate of $a_n$ will be specified in Theorem \ref{th2}. Meanwhile, If Eq. \eqref{eq5} has multiple solutions, only a sequence of consistent estimator $\widehat{\btheta}_n$ is considered. A sequence $\widehat{\btheta}_n$ is said to be a consistent sequence, if $\widehat{\bbeta}_n-\bbeta_{n0} \to \0$ and $\underset{x}{\sup}| \bB^\top(x)\widehat{\balpha}_n-g_0(x) |\to\0$ in probability as $n\to \infty$.
In what follows, we establish the asymptotic properties of the estimators of the parametric and nonparametric components for the GAMM under the proposed PGEE. In our asymptotic study, we assume the
number of subjects $n$ goes to infinity, $\phi=1$ and $n_i = m <\infty$. Extension of the methodology to the cases of unequal $n_i$ is straightforward. We vary the dimension of $\bA_i$ and replace $\widehat{\bR}$ by $\widehat{\bR}_i$, which is the $n_i \times n_i$ matrix using the specified working correlation structure and the corresponding initial parameter $\brho$ estimator. For simplicity, we further assume that the response, the functional, random and scalar predictors have been centered to mean zero for a total of $N=\sum_{i=1}^{n_i}$ observation. Without loss of generality, we denote $g(\mu_{ij})=\bX_{i(1)}\bbeta_{n1}+\bX_{i(2)}\bbeta_{n2}+\sum_{k=1}^{s_n}g_{ik}(x_{ijk})+\sum_{k=s_n+1}^{p_n}g_{ik}(x_{ijk})+\bz_{ij}^\top\bu_i$, where $\bbeta_{n1}$ is a $q_n\times 1$ vector corresponding to significant covariate effects $\bX_{i(1)}$ and $\bbeta_{n2}$ is a $(p_n - q_n)\times1$ vector of zeros, and $g_{ik}(.)=0$ almost surely for $k= s_n +
1,\ldots,r_n$, and $s_n$ is the total number of nonzero function components. Therefore $\lbrace\balpha_{nk}\rbrace_{k=1}^{s_n}$ are nonzero vectors associated to relevant functions $\lbrace g_{ik}(.)\rbrace_{k=1}^{s_n}$ and the $\lbrace\balpha_{nk}\rbrace_{k=s_n+1}^{r_n}$ are vectors of zero corresponding to redundant functions $\lbrace g_{ik}(.)\rbrace_{k=s_n+1}^{r_n}$. For technical convenience let  $\btheta_{n0}=(\btheta_{n01}^\top,\btheta_{n02}^\top)^\top$ be the true parameters, where $\btheta_{n01}=(\bbeta_{n01}^\top,\balpha_{n01}^\top,\ldots, \balpha_{n0s_n}^\top)^\top$ is $(s=q_n+\sum_{k=1}^{s_n}h_{nk})$-dimensional vector of that the elements are all nonzero and $\btheta_{n02}=(\bbeta_{n02}^\top,\balpha_{n0s_n+1}^\top,\ldots, \balpha_{n0r_n}^\top)^\top=\0$ is a vector of dimension $(p_n-q_n)+\sum_{k=s_n+1}^{r_n}h_{nk}$.
Consequently, estimated values and the design matrix is re-partitioned as $\widehat{\btheta}_n=(\widehat{\btheta}_{n1}^\top,\widehat{\btheta}_{n2}^\top)^\top$, and $\bD_{i}=\big(\bD_{i(1)}^\top, \bD_{i(2)}^\top\big)^\top$ which $\widehat{\btheta}_{n1}=(\widehat{\bbeta}_{n1}^\top,\widehat{\balpha}_{n1}^\top,\ldots, \widehat{\balpha}_{ns_n}^\top)^\top$,
$\widehat{\btheta}_{n2}=(\widehat{\bbeta}_{n2}^\top,\widehat{\balpha}_{ns_n+1}^\top,\ldots, \widehat{\balpha}_{nr_n}^\top)^\top$,
$\bD_{i(1)}=\big(\bX_{i(1)}^\top, \bB_1^\top,\ldots,\bB^\top_{s_n}\big)^\top$, and $\bD_{i(2)}=\big(\bX_{i(1)}^\top, \bB_{s_n+1}^\top,\ldots,\bB_{r_n}^\top\big)^\top$. Throughout, we need some regularity conditions and to save space we managed to
put them in a separate supplementary file (SF) and refer here as (A.1)-(A.7). Further, some lemmas are also used that we put them in the SF.
The following theorems establish the selection and estimation consistency for both the functional regression coefficients and regression parameters. Proofs of the following results can be found in the SF, also.

\begin{thm}
	Assume conditions (A.1)--(A.7) and that, the number of knots $L_n=O_p(n^{1/(2r+1)})$ and the tuning parameter $\lambda_n \to 0$. Then there exists a solution of $\bU_n(\btheta_n)= o(a_n)$, such that
	\begin{eqnarray*}
		(i)&&\Vert \widehat{\btheta}_n-\btheta_{n0} \Vert = O_p(\sqrt{p_n/n}),\cr
		(ii)&& \max_{1\leq k\leq p_n}\frac{1}{n}\sum_{i=1}^n\sum_{j=1}^{n_i}\big(\widehat{g}_{ik}(x_{ijk})-g_{0ik}(x_{ijk})\big)^2=O_p(n^{-2r/(2r+1)}).
	\end{eqnarray*}
	\label{th2}
\end{thm}

\begin{thm}
	Assume conditions (A.1)--(A.7), and that the tuning parameter $\lambda_n \to 0$, and $\lambda_{n} n^{r/(2r+1)} \to +\infty$. Then we have
	
	\begin{eqnarray*}
		(i)&& \P_n(\widehat{g}_k(.)=\0)\to 1,\quad \forall~ k=s_n+1,\ldots,p_n,\cr
		(ii)&& \P_n(\widehat{\bbeta}_{n2}=\0)\to 1,\cr
		(iii)&& \bxi_n^\top\overline{\bM}_n^{*^{-1/2}}(\bbeta_{n0})\overline{\bH}_n^*(\bbeta_{n0})(\widehat{\bbeta}_{n1}-\bbeta_{n01})\overset{\mathcal{D}}{\to} \hbox{N}_{p_n}(0,1),
	\end{eqnarray*}
	where
	\begin{eqnarray*}
		\overline{\bM}_n^{*}&=&\E_{u|y} \Big[\sum_{i=1}^n \bX_i^{*^\top}  \bA_i^{\frac{1}{2}}(\btheta_n , \bu_i ) \overline{\bR}^{-1}\bR_0 \overline{\bR}^{-1} \bA_i^{\frac{1}{2}}(\btheta_n , \bu_i )\bX_i^{*}\Big],\\
		\overline{\bH}_n^{*}&=&\E_{u|y} \Big[\sum_{i=1}^n \bX_i^{*^\top}  \bA_i^{\frac{1}{2}}(\btheta_n , \bu_i ) \overline{\bR}^{-1} \bA_i^{\frac{1}{2}}(\btheta_n , \bu_i )\bX_i^{*}\Big],
	\end{eqnarray*}
	$$\bX_i^{*}=(\bI-\bP)\bX_i,\quad \bP=\bB(\bB^\top\bOmega\bB)^{-1}\bB^\top\bOmega, \quad \bOmega=\hbox{diag}\lbrace \bOmega_i\rbrace$$
	and 
	$$\bOmega_i=\E_{u|y} \Big[\bA_i^{\frac{1}{2}}(\btheta_n , \bu_i ) \overline{\bR}^{-1} \bA_i^{\frac{1}{2}}(\btheta_n , \bu_i )\Big].$$
	\label{th3}
\end{thm}

\section{Numerical Studies}
In this section, in a series of numerical expriments, we assess the performance of our proposal.

\subsection{Simulations}
In this section, we conduct three simulation studies from the GAMM with both continuous and binary outcomes for 100 simulated data sets. For evaluating the estimation accuracy, we report the empirical mean square error (MSE), defined as $\sum_{r=1}^{100}\Vert\widehat{\bbeta}_n^{r}-\bbeta_{n0}\Vert/100$ where $\widehat{\bbeta}_n^r$ is the estimator of $\bbeta_{n0}$ obtained using the $r$th generated data set. We evaluate the total averaged integrated squared error (TAISE) to assess estimation efficiency of the functional part. Let $\widehat{g}^r$ be the estimator of a nonparametric function, $g$, in the
$r$th replication and $\lbrace x_m \rbrace_{m=1}^{n_{grid}}$ be the grid points where $\widehat{g}^r$ is evaluated. We define
$$
\rm{AISE}(\widehat{g})=\frac{1}{100}\sum_{r=1}^{100}\frac{1}{n_{grid}}\sum_{m=1}^{n_{grid}}\lbrace \widehat{g}^r(x_m)-g(x_m) \rbrace^2
$$
and $\rm{TAISE}=\sum_{k=1}^{p_n}\rm{AISE}(\widehat{g}_k)$. 
The performance in variable selection is gauged by 'FZf', number of false zero functional predictors; 'FNf', number of false nonzero functional predictors; 'FZs', number of false zero scalar covariates; 'FNs', number of false nonzero scalar covariates. To present a more comprehensive picture, we also use other criteria for variable selection performance evaluation. 'U.fit', 'C.fit', and 'O.fit' give the percentage of under fitting, correct fitting and over fitting  from 100 replications, respectively.

\subsubsection{Example 1: Continuous response}
In this example, the continuous responses $y_{ij}$ conditional on independent random effect $u_i\sim \mathcal{N}(0,0.5)$ are generated
from
\begin{eqnarray}\label{eq9}
y_{ij}|u_i=\sum_{k=1}^{r_n}x_{ijk}\beta_k+\sum_{k=1}^{p_n}g_{k}(x_{ijk})+u_i+\epsilon_{ij}, \quad i=1,\ldots,p_n;\qquad j=1,\ldots,5,
\end{eqnarray}
where $p_n=10$ and the number of sample size is $n=100, 250$ or $500$. The true regression
coefficients are $\bbeta=(-1,-1,2,0,\ldots,0)$ and the additive functions are
$$g_1(x_{ij1})=(2x_{ij1}-1)^2, \quad g_2(x_{ij2})=8(x_{ij2}-0.5)^3, \quad g_2(x_{ij2})=\sin(2\pi x_{ij3})$$
and $g_k(x_{ijk})=0$ for $k = 4,\ldots,10$. Thus the last seven coefficients and functional variables in this model are redundant and do not contribute to the model. The covariates $\bX_{ij}= (x_{ij1},\ldots,x_{ijp_n})^\top$ are generated independently from Uniform $(\lbrack 0,1 \rbrack^{10})$. The error $\bepsilon_i=(\epsilon_{i1}, \ldots,\epsilon_{i5})^\top$ follows a multi-5-variate normal distribution with mean $0$, a common marginal variance $\sigma^2=1$, and an AR-1 correlation structure with correlation $\rho= 0.7$.
We apply the PGEE with SCAD penalty and fit the
linear splines ($d$ = 1) and cubic splines ($d$ = 3). To illustrate the effect incorporating within-cluster correlation on estimation efficiency, we compare the estimation efficiency of using basis matrices from different working correlation structures: exchangeable (EX), AR-1, and independent (IND). In addition, we compare the PGEE approach with the GEE estimations of a full model (FULL) and an oracle model (ORACLE). Here, the full model consists of all 10 variables and the oracle model contains only the first three relevant variables. The results of this example, where considered moderate dimension of covariates, are described in the supplementary file.

To assess our method in more challenging cases for high-dimensional data, we consider a model with the dimension of functional components $p_n=100$ in \eqref{eq9}. However, only the first three variables are relevant and take the same functional forms as in Example 1. We consider the model \eqref{eq9} with moderate number of $n=200$ and $n_i=5$, with errors $\lbrace \bepsilon_i =(\epsilon_{i1} ,\ldots,\epsilon_{i5})^\top\rbrace_{i=1}^{200}$
generated independently from a multivariate normal distribution with mean $0$, and an AR-1 correlation structure with $\corr(\epsilon_{ij},\epsilon_{ij^{'}} ) = 0.7^{|j-j^{'}|}$ for $1 \leq j ,j^{'} \leq 5$.
We apply the  GEE to estimate the full and oracle models with working independent, exchangeable, or AR-1 working correlation. For variable selection, we consider the PGEE with SCAD penalty with basis matrices from IND, EX, or AR-1 working correlations. Table \ref{tab3} reports MSEs and TAISEs for FULL, ORACLE, and SCAD and variable selection results on correct, overfit, and underfit percentages of the SCAD approach for three IND, EX and AR-1 working correlations. Table \ref{tab3} clearly indicates that the improvement from incorporating within-cluster correlation is very significant. In particular, the estimation procedures with a correctly specified AR-1 structure always give smaller MSEs and TAISEs than those with a misspecified EX or IND working correlation. Also, the efficiency gained by incorporating correlation could be increased
in the cubic splines approach. For variable selection, the SCAD in both cases of linear and cubic, with an AR-1 working correlation also performs noticeably better than the one with EX or IND working correlation. Furthermore, Table \ref{tab3} also shows that the SCAD procedure dramatically improves the estimation accuracy for this high-dimensional case, with smaller MSEs and TAISEs from the FULL model.

\begin{table}[!ht]
	\caption{Example 2: Continuous response and high dimension of covariates. The MSEs (TAISEs in parentheses), and percentages of correct fitting (C.fit), underfitting (U.fit), and overfitting (O.fit) from 100 replications}
	\adjustbox{ max height = \dimexpr\textheight-5.7cm\relax, max width=\textwidth}{%
		\begin{tabular}{cccccccccccc}\\ \hline
			&&SCAD &ORACLE &FULL &FZs &FNs &FZf &FNf &U.fit &C.fit &O.fit\\
			\multirow{3}*{linear spline}
			&AR-1 &0.018(0.073) &0.027(0.659) &1.662(0.576) &0.02 &0.00 &0.00 &1.00 &0.02 &0.98 &0.00\\
			&EX   &0.031(0.113) &0.056(1.003) &1.056(0.837) &0.01 &2.22 &0.00 &0.00 &0.01 &0.94 &0.05\\
			&IND  &0.029(0.197) &0.078(1.002) &3.206(4.991) &0.18 &0.02 &0.00 &0.00 &0.17 &0.81 &0.02\\
			\multirow{3}*{cubic spline}
			&AR-1 &0.014(0.062) &0.013(0.056) &2.273(0.081) &0.10 &0.00 &0.00 &0.00 &0.07 &0.93 &0.00\\
			&EX   &0.016(0.088) &0.014(0.152) &2.571(0.142) &0.15 &0.00 &0.00 &1.00 &0.12 &0.88 &0.00\\
			&IND  &0.011(0.147) &0.018(0.076) &3.017(4.617) &0.27 &0.35 &0.00 &1.00 &0.26 &0.68 &0.06\\  \hline
		\end{tabular}
	}
	\label{tab3}
\end{table}

\subsubsection{Example 3: Binary response}
To assess the performance of our method for binary outcomes, we generate a moderate random sample of 200 subjects in each simulation run. Within each subject, binary responses $\lbrace y_{ij}\rbrace_{j=1}^{200}$ are generated from a marginal logit model
$$\logit\P(y_{ij}=1|\bX_{ij}=\bX_{ij})=\sum_{k=1}^{10}x_{ijk}\beta_k+\sum_{k=1}^{10}g_{k}(x_{ijk})+u_i$$
where $g_1(x_{ij1})=\lbrack \exp(x+1)-\big(\exp(2)-\exp(1)\big)/16 \rbrack$,
$g_2(x_{ij2})=\cos(2\pi x_{ij2})/4$
$g_3(x_{ij3})=x_{ij3}(1-x_{ij3})-1/6$
$g_4(x_{ij4})=2(x_{ij4}-0.5)^3$ and
the remaining $6$ covariates are null variables with $g_4(x_{ij4})=0$ for $k=5,\ldots,10$. The covariates $\bX_{ij}= (x_{ij1},\ldots,x_{ijp_n})^\top$ are generated independently from Uniform $(\lbrack 0,1 \rbrack^{10})$. We generate correlated binary responses with AR-1 correlation structure
with a correlation parameter of $0.5$.
We applied the PGEE with a SCAD penalty for the variable selection, and the GEE for estimation of the full and oracle models. To illustrate how different working correlations could affect our estimation and variable selection results, we consider EX and IND structures, in addition to the true AR-1 correlation structure.
Table \ref{tab4} gives the TAISEs for the SCAD, ORACLE, and
FULL models with three different working correlations. Similar to the previous continuous simulation studies, MSEs and TAISEs calculated based on the both linear and cubic SCAD approach are also shown to be close to the MSEs and TAISEs from ORACLE, and much smaller than those from the FULL model.

For all three aformentioned study, the amounts of FZs, FNs, FZf, FNf, U.fit, C.fit and O.fit shows that the SCAD approach can gain significant estimation accuracy by effectively removing the zero component variables. Overall, the SCAD procedures work reasonably well, and the SCAD with EX and AR-1 working correlation structures provides better variable selection results than the SCAD with IND working structure. In addition, the SCAD approach with AR-1 correlation performs the best in selecting non-zero component variables.

\begin{table}[!ht]
	\caption{Example 3: Binary response. The MSEs (TAISEs in parentheses), and percentages of correct fitting (C.fit), underfitting (U.fit), and overfitting (O.fit) from 100 replications}	
\adjustbox{ max height = \dimexpr\textheight-5.7cm\relax, max width=\textwidth}{%
		\begin{tabular}{cccccccccccc}\\ \hline
			&&SCAD &ORACLE &FULL &FZs &FNs &FZf &FNf &U.fit &C.fit &O.fit\\
			\multirow{3}*{linear spline}
			&AR-1 &0.443(0.249) &0.209(0.000)  &0.456(0.972)  &0.00 &0.00  &0.00 &0.00  &0.000  &1.00  &0.00\\
			&EX   &0.451(0.262) &0.207(0.000) &0.456(0.988) &0.00 &0.00  &0.00 &0.00 &0.00 &1.00 &0.00\\
			&IND  &0.447(0.236) &0.209(0.000)  &0.455(0.980)   &0.01  &0.00 &0.00 &0.00 &0.01 &0.99 &0.00\\ 
			\multirow{3}*{cubic spline}
			&AR-1 &0.474(0.000) &0.257(0.000) &1.899(1.759) &0.14 &0.00& 0.00& 2.00& 0.14& 0.86& 0.00\\
			&EX   &0.478(0.000) &0.257(0.000) &1.903(1.761) &0.15 &0.00 &0.00 &2.00 &0.15 &0.85 &0.00\\
			&IND  &0.477(0.000) &2.579(0.000)&1.882(1.746)  &0.15 &0.00 &0.00 &2.00 &0.15 &0.85 &0.00\\ \hline
	\end{tabular}}
	\label{tab4}
\end{table}

\subsection{Benchmark analysis}
In this subsection, we analyse the data from the Multi-Center AIDS Cohort study. The data set contains the human immunodeficiency virus, HIV, status of 283 homesexual men infected with HIV during the follow-up period between 1984 to 1991. All individual were scheduled to have their measurment made during semi annual visit. Here $t_{ij}, i=1, \ldots, n, j=1, \ldots, n_i$ denote the time length in years between seroconversion and the $j$th measurment of the $i$th individual after the infection. Xue and Zhu (2007) analyzed the data set using partial linear models. The primary interest was to describe the trend of the mean CD4 percentage depletion over time and to evaluate the effect of cigarette smoking, pre-HIV infection CD4 percentage, and age at infection on the mean CD4 cell percentage after the infection. We take four covariates for this study and construct an additive model by including all these covariates in both parametric and functional parts. We apply the PGEE with SCAD penalty and fit the linear splines ($d$ = 1) and cubic splines ($d$ = 3). In addition, we compare the PGEE approach with the GEE estimations of a full model. We use the standard errors (SE) were all calculated using the leave-one-out cross validation method. To best identify a model supported by the data, mean squared estimation errors (MSEE) and mean squared prediction errors (MSPE) were calculated for each method. 

\begin{figure}[!ht]
	\begin{center}
		\includegraphics[width=12cm, height=6cm]{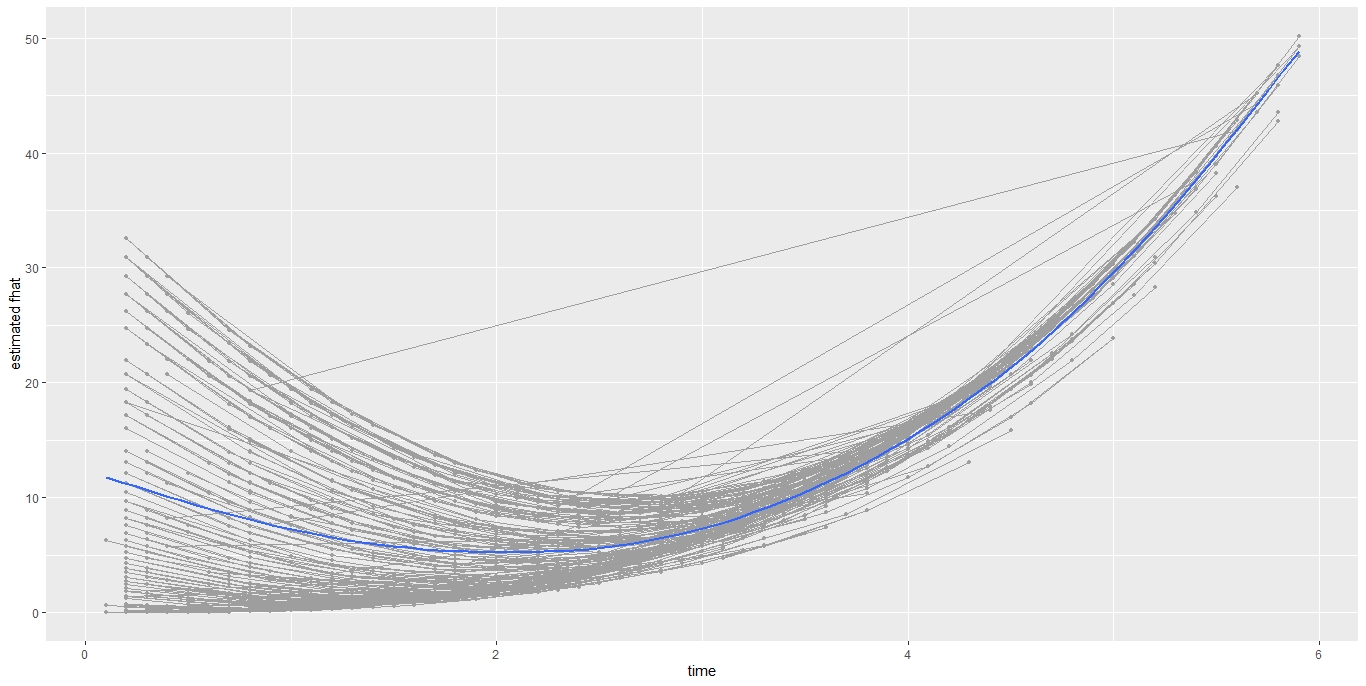}
		\caption{ The estimated component functions ($\widehat{f}(t)$) from the SCAD procedure with cubic spline.}\label{fig1}
	\end{center}
\end{figure}

Table \ref{tab5} reports the results. Two variable selection methods resulted in models with smaller SE, MSEE, and better predictions than the full model, which indicates that redundant variables exist in data sets. Cubic spline gave parsimonious models with the best prediction performance. Linear spline select the all nonparametric functions as zero, but cubic spline and FULL models select time variable is nonparametric. As shown in figure \ref{fig1}, we see that mean baseline CD4 perecentage of the population (blue line) and all subjects (gray lines) depreters rather quickly at the begining of HIV infection, but the CD4 perecentage appears to be incresing two years after the infection. Maybe taking medication has caused such behavior.

\begin{table}[!ht]
	\centering
	\caption{CD4 data examples: estimation, prediction and variable selection results.}{%
		\begin{tabular}{ccccccc}\\ \hline 
			&\multicolumn{2}{c}{linear spline} &\multicolumn{2}{c}{cubic spline} &\multicolumn{2}{c}{FULL}\\
			variabels &estimete &SD &estimete &SD &estimete &SD\\
			time    &0.000  &0.000 &-6.247 &$1.4\times 10^{-4}$ &-0.448 & 1.069 \\
			pre-CD4 &-0.342 &0.114 &0.175  &$3.5\times 10^{-4}$ &-0.457 &0.095\\
			age     &0.556  &0.126 &1.576  &$5.3\times 10^{-4}$ &0.359  &0.113\\
			smoke   &0.000  &0.000 &-0.953 &$4.2\times 10^{-6}$ &-0.515 &1.512\\\hline
			MSEE    &\multicolumn{2}{c}{0.240}  &\multicolumn{2}{c}{0.049}  &\multicolumn{2}{c}{13.623}\\
			MSPE    &\multicolumn{2}{c}{10.101} &\multicolumn{2}{c}{9.422} &\multicolumn{2}{c}{11.534} \\ \hline
	\end{tabular}}
	\label{tab5}
\end{table}

\subsection{Yeast Cell-Cycle Gene Expression data analysis}
A yeast cell-cycle gene expression data collected in the CDC15 experiment of  Spellman et al. (1998) where genome-wide mRNA levels of 6178 yeast ORFs (abbreviation for open reading frames, which are DNA sequences that can determine which amino acids will be encoded by a gene) at 7 minute intervals for 119 minutes, which covers two cell-cycle periods for a total of 18 time points, measured. The cell cycle is a tightly regulated life process where cells grow, replicate their DNA, segregate their chromosomes, and divide into as many daughter cells as the environment allows. The cell-cycle process is commonly divided into M/G1-G1-S-G2-M stages. Refer to Wang et al. (2012), for more detailed description of this data set.
Transcription factors (TFs) have been observed to play critical roles in gene expression regulation. A TF (sometimes called a sequence-specific DNA-binding factor) is a protein that binds to specific DNA sequences, thereby controlling the flow (or transcription) of genetic information from DNA to mRNA. To better understand the phenomenon underlying cell-cycle process, it is important to identify TFs that regulate the gene expression levels of cell cycle-regulated genes. It is not clear where these TFs regulate all cell cycle genes, however. We applied our methods to identify the key TFs. The dataset that we use present a subset of 283 cell-cycled-regularized genes observed over 4 time points at G1 stage. The response variable $Y_{ij}$ is the log-transformed gene expression level of gene $i$ measured at time point $j$, for $i = 1,\ldots, 283$.  We use the following GAMM model
\begin{eqnarray*}
	y_{ij}= \sum_{k=1}^{96} x_{ijk}+\sum_{k=1}^{96}g_{ik}(x_{ijk})+u_i,
\end{eqnarray*}
where the covariates $x_{ijk}$ , $k = 1,\ldots, 96$, is the matching score of the binding probability of the $k$th TF on the promoter region of the $i$th gene. The binding probability is computed using a mixture modeling approach based on data from a ChIP binding experiment; see Wang et al. (2007) for details. Covariates $x_{ijk}$ is standardized to have mean zero, and $u_i$ is the random intercept. Our goal is to identify the TFs that might be related to the expression patterns of these 283 cellcycle-regulated genes. Therefore we apply a penalization procedure by the proposal P-GAMM and fit the linear splines ($d$ = 1) and cubic splines ($d$ = 3), and also use the full model for comparing. Table \ref{tab6} reports the estimation and prediction performance when linear and cubic splines and full model are adopted. For stage G1, MBP1, SWI4, and SWI6 are three TFs that have been proved important in the aforementioned biological experiments and our analysis reveals that they have been selected by the penalization method with linear and cubic splines. Among this important TFs, linear spline select the all nonparametric effects as zero, but cubic spline identify that MBP1 and SWI4 have nonparametric effects too. Our proposal in both cases of linear and cubic give smaller MSEE and MSPE then the FULL model, where show that the penalized GAMM improves the estimation and prediction performance. Prediction accuracy are similar in both linear and cubic splines, but linear spline perform more efficient in term of MSEE. 

\begin{table}[!ht]
	\centering
	\caption{Yeast Cell-Cycle Gene Expression data: estimation and prediction prediction performance.}{%
		\begin{tabular}{cccc}\\\hline
			&linear spline &cubic spline &FULL\\
			MSEE &0.002 &0.176 &4.141 \\
			MSPE &0.351 &0.339 &0.463 \\ \hline
	\end{tabular}}
	\label{tab6}
\end{table}

\section{Concluding Remarks}
In general, when the number of covariates is large, identifying the exact underlying model is a challenging task, in particular when some of the nonzero signals are relatively weak. Here, we consider the GAMM when longitudinal responses collected in a high dimensional regim. In our setting, there is no specified likelihood function for the GAM, because the outcomes could be nonnormal and discrete, which makes estimation and model selection very challenging problems. Also, we found that it may be quite computationally intensive in high-dimensional variable selection settings, because the dimension of the parameters in the nonparametric forms increases significantly compared with parametric model selection settings. We approximated the additive functional components based on polynomial B-spline smoothing and propose doubel penalized method where variable selection and estimation of the both parametric and nonparametric components obtained, and were able to select the functional components groupwise. The procedure involved the specification of the posterior distribution of the random effects, which cannot be evaluated in a closed form. We used a Metropolis algorithm which does not require the specification of the posterior distribution. To implement the procedure in practice, a computationally flexible iterative algorithm developed. We established an asymptotic property with consistency for the fixed effect covariates and nonparametric components, which achieves the optimal rate of convergence. In addition, the proposed model selection strategy was able to select the correct GAMM consistently. That is, with probability approaching to 1, the estimators for the zero function components converge to 0 almost surely. We illustrated our method using numerical studies with both continuous and binary responses, along with real data applications. Results demonstrated that the proposal works well, and can correctly select the nonzero fixed effect covariates and functional components with probability tending to one as the sample size goes to infinity. 

In the proposed model same set of covariates is included in both linear and nonparametric components. The statistical inference of these model can be more reliable by identifying that the parametric and nonparametric parts are known in advance. However, such prior information is usually unavailable, especially when the number of covariates is large. Therefore, it is of great interest to develop some efficient methods to distinguish parametric components from nonparametric ones. For future work, one can consider a procedure, in the context of high dimensional GAMM, which aims to reduce the size of covariates vector and distinguish linear and nonlinear effects among nonzero components similar to those in \citet{Kazemi:et:al:2019}.

\section*{Acknowledgments}
Mohammad Arashi's work is based upon research funded by the Iran National Science Foundation (INSF) grant No. 4015320.






\end{document}